\documentclass[aps,floatfix,onecolumn,a4paper,showpacs, nofootinbib,superscriptaddress,10pt]{revtex4}
\usepackage{amssymb,amsmath}
\usepackage[dvips]{graphics,color}
\usepackage{epsfig}\usepackage{float}\usepackage{float,upgreek,yfonts}
\usepackage{graphicx}
\newcommand{\ba}{\begin{eqnarray}}
\newcommand{\ea}{\end{eqnarray}}
\newcommand{\bege}{\begin{equation}}
\newcommand{\enge}{\end{equation}}

\newcommand{\beq}{\begin{eqnarray}}
\newcommand{\benu}{\begin{enumerate}}
\newcommand{\enu}{\end{enumerate}}
\newcommand{\eeq}{\end{eqnarray}}

\usepackage[colorlinks,hyperindex]{hyperref}
\usepackage{xcolor}
\definecolor{green1}{RGB}{0,128,0} 
\hypersetup{hidelinks,backref=true,pagebackref=true,hyperindex=true,colorlinks=true,breaklinks=true,urlcolor= blue}
\hypersetup{%
  colorlinks = true,
  linkcolor  = blue,
  citecolor = green1,
}
\usepackage{hyperref,color,xcolor}
\hypersetup{hidelinks,hyperindex=true,colorlinks=true,breaklinks=true,urlcolor= blue}
\hypersetup{%
  colorlinks = true,
  linkcolor  = blue
}

\begin{document}

\title{Spinor symmetries and underlying properties}

\author{J. M. Hoff da Silva}
\email{julio.hoff@unesp.br} \affiliation{Departamento de F\'{\i}sica e Qu\'{\i}mica, Universidade Estadual Paulista, UNESP, Guaratinguet\'{a}, SP, Brazil}

\author{R. T. Cavalcanti}
\email{rogerio.cavalcanti@unesp.br} \affiliation{Departamento de F\'{\i}sica e Qu\'{\i}mica, Universidade Estadual Paulista, UNESP, Guaratinguet\'{a}, SP, Brazil}

\author{D. Beghetto}
\email{dbeghetto@feg.unesp.br} \affiliation{Departamento de F\'{\i}sica e Qu\'{\i}mica, Universidade Estadual Paulista, UNESP, Guaratinguet\'{a}, SP, Brazil}

\author{R. da Rocha}
\email{roldao.rocha@ufabc.edu.br}
\affiliation{Federal University of ABC, Center of Mathematics, 09210-580, Santo Andr\'e, Brazil}
\affiliation{International Centre for Theoretical Physics, Strada Costiera 11, 34151 Trieste, Italy}


\begin{abstract}
By exploring a spinor space whose elements carry a spin 1/2 representation of the Lorentz group and satisfy the the Fierz-Pauli-Kofink identities we show that certain symmetries operations form a Lie group. Moreover, we discuss the reflex of the Dirac dynamics in the spinor space. In particular, we show that the usual dynamics for massless spinors in the spacetime is related to an incompressible fluid behavior in the spinor space. 
\end{abstract}
\maketitle

\section{Introduction}

Spinors have constituted a comprehensive mathematical object of study, with 
a variety of applications in Physics. In particular, spinors are the main ingredient
in the description of fermionic particles, that encode ordinary matter in the Universe. 
Fundamentally constructed upon the Lorentz group, a lot have been looked at 
the spacetime symmetries, underlying the Lounesto classification, rather than the symmetries of the spinor space itself. The Lounesto classification of spinor fields
allocates classical spinors into six disjoint classes of regular and singular spinors.
Regular spinors encompass Dirac spinors, and the singular ones, constituting flag-dipole, flagpole and dipole structures, comprehend the 
Majorana and Weyl spinors. Refs. \cite{Cavalcanti:2014wia,Fabbri:2016msm} constructed a 
reciprocal classification for spinors, including gauge field theoretical aspects. Astonishingly, spinors satisfying the Dirac equation 
in several backgrounds have been found into five out of the six Lounesto's classes.
However, the subclasses of spinors, in each spinor class, whose equations of motion have been already stablished, are not fully determined. In other words, 
each spinor class has subclasses with precise dynamics, but the problem of 
categorizing the dynamics of all spinors in each class is intricate. Several approaches 
were scrutinized in Refs. \cite{Bonora:2017oyb,Fabbri:2017lvu,Fabbri:2016msm,daRocha:2016bil,daRocha:2005ti,daRocha:2013qhu,Ablamowicz:2014rpa,5}. However, for example the fifth class of singular spinors
has at least three subclasses:  neutral spinors satisfying Majorana equation, 
 eigenspinors of the charge conjugation operator with dual helicity, satisfying 
 Elko coupled first order equations of motion, and also charged spinors satisfying the Dirac equation that induce an underlying fluid flow structure in some background spacetimes \cite{daRocha:2016bil}. The Lounesto spinor classification, encompassing classes of charged and neutral spinors under the U(1) gauge symmetry, was extended in Ref. \cite{Fabbri:2017lvu} to non-Abelian gauge symmetries. A second-quantized field theoretical approach \cite{Bonora:2017oyb} poses a similar  classification in the framework of second quantization. Unexpected tensorial objetcs emerging from the spinor dynamics were also found in Ref. \cite{Fabbri:2018crr}.   Therefore, it is clear that the question  regarding the dynamics and kinematics of all spinors in Lounesto classification lacks still. 

Within this motivation, a hybrid paradigm, uniting the symmetries on the spinor space and 
the spinors as representatives of the spinor classes in Lounesto classification, was previously proposed \cite{propo}. The ideia was to use this spinor representation space to envisage physical characteristics as an output of geometric, algebraic and topological properties of the constructed space, $\Sigma$, by exploring the point of view of spinors completely characterized by its bilinear covariants. In this work we continue exploring such a space, this time further exploring symmetries acting upon spinors themselves. Naturally, these symmetries should preserve each one of the spinor classes in Lounesto classification. Starting from its definition we investigate symmetries properties in the spinor space. It is shown that (invertible) symmetries transformations are rescaling for every bilinear covariants components constituting a subgroup of $GL(4,\mathbb{C})$. Also, the possibility of projective representations is explored, where we highlight an algebraic parallel of a superselection rule. These results are presented in Sec. III which is preceded by a review about the Lounesto spinor classification.  

Section IV is reserved to the investigation of the usual Dirac dynamics in the spinor space $\Sigma$. Assuming the existence of a homomorphism between $\Sigma$ and an open set of ${\bf P}_{Spin_{1,3}^e}\times_\tau \mathbb{C}^4$ (see further specifications of this bundle in Sec. II) it is shown, under certain general conditions, that for massless spinors an analog of the Liouville theorem may be set. This section is finished contrasting such a result for the case of exotic spinors. In the final section we conclude.

\section{Lounesto classification}
\label{II}
  Let $M$ denote the Minkowski spacetime. Spinors are objects in the spinor bundle,  ${\bf P}_{Spin_{1,3}^e}\times_\tau \mathbb{C}^4$ associated to $M$, carrying the so-called $\tau={\left(1/2, 0\right)}\oplus{\left(0, 1/2\right)}$ representations of the Lorentz group  \cite{1,2,3}.  Due to several applications, arbitrary bases $\{\upgamma^\mu\}\subset \Omega(M)=\oplus_{i=0}^4\Omega^i(M)$ of the exterior bundle may be adopted. The bilinear covariants are the following exterior bundle sections \cite{TAKA}, 
\begin{subequations}
\begin{eqnarray}
\textcolor{black}{\upsigma}&=&\textcolor{black}{\bar{\psi}\psi}\in\Upomega^0(M),
\label{upsig}\\
\mathbb{J}&=&\bar{\psi}\upgamma _{\mu }\psi\,\upgamma^\mu\in\Upomega^1(M),\label{jj}\\
\mathbb{S}&=&\frac{i}{2}\bar{\psi}[\upgamma _{\mu},\upgamma_{
\nu }]\psi\,\upgamma^\mu\wedge\upgamma^\nu\in\Upomega^2(M),\label{ss}\\
\mathbb{K}&=&\bar{\psi}\upgamma _{\mu }\upgamma^5\psi\,\upgamma^\mu\in\Upomega^3(M),\label{kk}\\
\upomega&=&{i\bar{\psi}\upgamma_{5}\psi\,,}\in\Upomega^4(M).\label{upom}
\end{eqnarray}
\end{subequations}
The generators $\{\upgamma^\mu\}$ also satisfy the Clifford-Dirac algebra,  $\upgamma_{\mu }\upgamma _{\nu
}+\upgamma _{\nu }\upgamma_{\mu }=2\eta_{\mu \nu }\mathbf{1}$. Besides, $\upgamma_5=i\upgamma_0\upgamma_1\upgamma_2\upgamma_3$. The spinor conjugation is denoted by $\bar\psi=\psi^\dagger\upgamma_0$. 

 Lounesto classification \cite{4,5} allocates spinors into classes according to their bilinear covariants, categorizing and organizing the physically relevant spinorial space which can, therefore, be faced as composed by these six distinct pieces. In fact, the Lounesto classification split off the following:
\begin{subequations}
\begin{eqnarray}
&&(1)\;\;\;\mathbb{K}\neq 0, \;\;\;\mathbb{S}\neq0,\;\;\;\upomega\neq0,\;\;\; \upsigma\neq0,\;\;\text{}
\label{tipo1}\\
&&(2)\;\;\;\mathbb{K}\neq 0, \;\;\;\mathbb{S}\neq0,\;\;\;\upomega\neq0,\;\;\;  \upsigma=0,\;\;\label{tipo2}\\   
&&(3)\;\;\;\mathbb{K}\neq 0, \;\;\;\mathbb{S}\neq0,\;\;\;\upomega=0,\;\;\;  \upsigma\neq0,\label{tipo3}\\  
&&(4)\;\;\;\mathbb{K}\neq 0, \;\;\;\mathbb{S}\neq0,\;\;\;\upomega=0=\upsigma,  \;\;\text{}\quad\qquad\label{tipo4}\\
&&(5) \;\;\;\mathbb{K}=0, \;\;\;\mathbb{S}\neq0,\;\;\;\upomega=0=\upsigma,\text{}\quad\qquad\label{tipo5}\\
&&(6)\;\;\;\mathbb{K}\neq0, \;\;\;\mathbb{S}=0,\;\;\;\upomega=0=\upsigma.\;\;\;\text{}\quad\qquad\label{tipo6}
\end{eqnarray}
\end{subequations}
Some physically important observables are interpreted as usual. For instance, the scalar $\sigma$ is the mass term in Lagrangians, the pseudoscalar $\omega$ bilinears can reveal CP violations,  whereas the current is given by $\mathbb{J}$. Some regular spinors and most of the singular ones in the above classes are not supported by the same physical interpretation given to the electron. Besides, particular subclasses of the Lounesto's classification satisfy the Fierz--Pauli--Kofink (FPK) relations 
\cite{4,TAKA}: 
\begin{subequations}
	\begin{eqnarray}\label{fifi1}
	-\upomega S_{\mu\nu}-\frac{\upsigma}{2}\epsilon_{\mu\nu\alpha\beta}S^{\alpha\beta}&=&J_\mu {K}_\nu-{K}_\mu J_\nu ,
	\\\eta_{\alpha\beta}(J^\alpha J^\beta+K^\alpha K^\beta)&=&0=\eta_{\alpha\beta}J^\alpha K^\beta,\label{fifi2}\\ \eta_{\alpha\beta}J^\alpha J^\beta&=&\upsigma^{2}+\upomega^{2}\,.\label{fifi3}  
	\end{eqnarray}
\end{subequations}
\noindent  
We remark, by passing, that spinors not obeying to the FPK equations, the so-called amorphous spinors \cite{Crumeyrolle:1990vj,Rodrigues:1996tj}, shall be treated in what follows as non-physical spinors.

\section{Symmetries in the spinorial space}
\label{II}

Let us denote a given bilinear by $\bar{\psi}\Gamma\psi$, for which $\Gamma$ is any element of the set $\{\mathbb{I},\gamma_5,\gamma_\mu,\gamma_5\gamma_\mu,\gamma_{\mu}\gamma_{\nu}\}$, where $\mathbb{I}$ stands for the identity matrix, and the spinor dual is the usual (Dirac) one. Lounesto classification depends whether a given subset of bilinear is null or not, respecting the FPK identities. The relevant aspect to be emphasized here is that, concerning classical spinors in physics, the Lounesto classification is based in the physical observables and, hence, the belonging to a given type is by itself a physical information. This remark motivates the following definition\footnote{For details in the spinorial space definition, see \cite{propo}. Here we just remark that this space comprises only spinors obeying the FPK identities.}.

\vspace{.2cm}
{\bf Definition:} {\it A symmetry in the physical spinor space is any linear or anti-linear operation preserving the spinor type.} 
\vspace{.2cm}

 Thus any phase multiplying a spinor is also a symmetry and one is facing a ray representation of spinors, very much like the use of Hilbert space vectors in quantum mechanics, representing physical states. Denoting by $\Sigma_i$ the part of the spinor space encompassing type-$i$ spinors $(i\in\{1, 2, \ldots,6\})$, with $\Sigma=\cup_{i=1}^6\Sigma_i$, a certain spinor $\psi$ is better characterized by an equivalence class representing its ray, denoted by $R$. Hence $\psi \in R \subset \Sigma_i \subset \Sigma$. 

We shall now explore symmetry transformations. Let $S$ be a transformation leading rays into symmetry-preserving rays, that is
\begin{eqnarray}
S_i: R\subset\Sigma_i&\to& R'\subset\Sigma_i, \forall i\in\{1, 2, \ldots,6\} \nonumber\\ 
\left[\psi\right] &\mapsto& [\psi ']=S_i[\psi], \label{pri}
\end{eqnarray} where $[\psi]$ denotes the equivalence class to which $\psi$ belongs, each class composed by spinors differing only by a phase. Therefore, if $\psi'$ is a spinor different of $\psi$, namely a modified spinor, a symmetry means $S_i(\Sigma_i) \subset \Sigma_i$. 
More explicitly, a symmetry should obey 
\begin{eqnarray}
[\bar{\psi}]\gamma^0S^\dagger\gamma^0\Gamma S[\psi]=\beta_{\Gamma} [\bar{\psi}]\Gamma[\psi],\label{se}
\end{eqnarray} where $\beta_{\Gamma} \in \mathbb{R}$ is the shift resulting from the transformation action. This shift will be non null, in general, and different from one\footnote{If $\beta_{\Gamma}=1$, for any $\Gamma$, the inversion theorem \cite{TAKA,TAKA1} yields necessarily $S=\mathbb{I}$.}. Depending on the bilinear dealt with, $\beta_\Gamma$ must be replaced by an array, or disposed into a matrix structure, albeit this is not important now. In the case of the $\Sigma_6$ space, the representation of a general symmetry $S_6$ can be straightforwardly displayed by a block diagonal matrix of one of the two following forms:\footnote{As the apparatus is representation-independent, we are adopting the Weyl representation for the $\gamma$ matrices.}
\begin{align}\label{stipo6}
 \left( \begin{array}{cc}
  A & \mathbb{O} \\ 
  \mathbb{O} & B
  \end{array} \right) \qquad \text{or} \qquad  \left( \begin{array}{cc}
  \mathbb{O} & A \\ 
  B & \mathbb{O}
  \end{array} \right),
  \end{align}  
where $A$ and $B$ are $2\times 2$ matrices with only $A$ or $B$ necessarily non-null. In fact, for a general singular spinor $\psi=(a,b,c,d)^\intercal$,  the transformation $S_i$, $i=4,5,6$, must preserve the algebraic relation $a=\frac{bcd^*}{\Vert c\Vert^2}$. On the other hand, the opposite relation  $a\neq \frac{bcd^*}{\Vert c\Vert^2}$, must be preserved for regular spinors \cite{Cavalcanti:2014wia}.

Being $\chi$ a mapping between $\Sigma$ and the dual space $\bar{\Sigma}$, which is defined in a quite similar manner to $\Sigma$, we will restrict our analysis to the case in which $\chi$ is one-to-one. The reason is simple: being $\chi$ not one-to-one, then to an element of a given $R$, say $e^{i\alpha}\psi$ with $\alpha \in \mathbb{R}$, it would correspond $e^{-i\beta}\bar{\psi}$ in $\bar{\Sigma}$, with $\beta$ possibly different of $\alpha$. Then $\bar{\psi}'\Gamma\psi'=e^{i(\alpha-\beta)}\bar{\psi}\Gamma\psi$, breaking the symmetry.  

 In the following we are interested in symmetries such that, for every $S$ leading from a ray to another one, there should exist an inverse mapping, $S^{-1}$, pulling the transformation back.  Besides, if $S_1$ transforms a ray $R$ into $R'$ and $S_2$ leads $R'$ into $R''$, then the acting of $S_1$ followed by $S_2$ should have the same effect of an unique transformation, say $S_3$, going directly from $R$ to $R''$. Taking all into account, provided associativity, symmetries transformations, if allowed, may form a group. We stress that the existence of symmetry transformations without inverse for all spinor types is, in principle, not forbidden. Not invertible symmetry transformations may also be physically relevant. However, we concentrate  in the invertible case, since we are interested in a possible group structure.\footnote{The particular case of type-6 spinors does not require the whole matrix $S_6$ being invertible for having the group structure. It is sufficient being the non-null block invertible.} 

\vspace{.2cm}
{\bf Lemma:} {\it The invertible symmetry transformations allowed are a simple rescaling, up to a sign, for all bilinear covariants components.}   
\vspace{.2cm} 

{\bf Proof:} Being the scalar and pseudo-scalar bilinear covariants non null, one may write 
\begin{eqnarray}
\gamma^0S^\dagger\gamma^0 S&=&\alpha\mathbb{I},\label{l21}\\
\gamma^0S^\dagger\gamma^0\gamma^5 S&=&\beta \gamma^5.\label{l22}
\end{eqnarray} These equations combine into 
\begin{equation}
\alpha S^{-1}\gamma^5 S=\beta \gamma^5. \label{l23}
\end{equation} Taking the determinant of both sides of (\ref{l23}) yields $\alpha=\pm \beta$. A similar reasoning may be straigtforwardly extended to all components of the bilinear covariants, covering all the possible types. Some remarks, nevertheless, are in order before concluding. First, the proportionality between a transformed tensorial bilinear may be performed by a tensorial quantity, as to allow -- respecting symmetry -- the vanishing of some given component and the raising of another one. In any case, the final value of the tensorial quantity components are subject to the analysis above. Finally, the possible change of sign must, obviously, respect the constraints coming from FPK identities.  \hspace{.2cm}  $\Box$   
\medbreak
With these results we are able to enunciate the next theorem.

\vspace{.2cm}
{\bf Theorem:} {\it The symmetry transformations allows the space of spinors to perform a subgroup of $GL(4,\mathbb{C})$.} 
\vspace{.2cm} 

{\bf Proof:} Let $\{X,Y,S,\ldots\}$ be a set of symmetry transformations for type-$1$ spinors in which every element belongs to $\mathbb{M}(4,\mathbb{C})$. Suppose $X$ and $Y$ both satisfying (\ref{se}) for, say, $\beta_{\Gamma X}$ and $\beta_{\Gamma Y}$, respectively. Hence 
\begin{eqnarray}
\gamma^0(XY)^\dagger\gamma^0\Gamma(XY),\label{pre}
\end{eqnarray} shall also satisfy (\ref{se}) for $\beta_{\Gamma }=\beta_{\Gamma X}\beta_{\Gamma Y}$. Besides, it is fairly simple to see that the inverse transformation respects $\gamma^0(S^{-1})^\dagger\gamma^0\Gamma S^{-1}=\beta^{-1}_{\Gamma }\Gamma$. Once again we remark that when necessary the proportionality, and its inverse, must be engendered by a tensorial object. \hspace{.2cm}  $\Box$
\medbreak

We finalize this section by stating some facts about the representation of the symmetry group found in the spinor space. As a matter of fact, while symmetry transformations act upon rays, the operators representing the above group transform spinors itself. In this regard, the representation will inherit most of the group properties. Denoting by $O(S)$ the operator representing the symmetry action in the spinor space, the resulting state $O(S_1)O(S_2)\psi$ differs from $O(S_1 S_2)\psi$, as usual, by a phase at most. This is, of course, the indication of a possible projective representation. At this point we have not enough information about the topology of the subgroup referred in the above theorem, although the elimination of its central charge seems to be reachable. Hence we willl postpone the elimination, so to speak, of the projective representation for the future. Instead we would like to point out an interesting peculiarity of the representation.  

When dealing with representation up to a phase $O(S_1)O(S_2)\psi_k=e^{\phi_k}O(S_1 S_2)\psi_k$ the usual approach to quantum states yields a phase that does not depend on the state (here evinced by the label $k$) upon which the operators act, exception made to forbidden states, whose existence is precluded by means of a superselection rule \cite{WIWI}. The general picture may be straightforwardly recalled as follows: taking the sum of two spinors, say $\psi_m$ and $\psi_n$, and representing the transformation we have $O(S_1)O(S_2)(\psi_m+\psi_n)=e^{i\phi_{mn}}O(S_1 S_2)(\psi_m+\psi_n)$. After working out the right-hand side and acting with $O^{-1}(S_1S_2)$, we are left with
\begin{equation}
e^{i\phi_m}\psi_m+e^{i\phi_n}\psi_n=e^{i\phi_{mn}}(\psi_m+\psi_n),\label{phase}
\end{equation} where $O$ is assumed unitary, for simplicity. Clearly, a solution for the above equation is $\phi_{mn}=\phi_m=\phi_n$ pointing to a phase independent to the state, but as symmetries transformations are allowed in this space, $\psi_m$ and $\psi_n$ may well be connected, and therefore it is hard to accept that the independence of the phases is reached by chance. In this regard, the very existence of the symmetry may be faced as the analogue of the superselection rule. As a final comment, we remark that the reasoning just outlined cannot be applied to type-$i$ spinors as a whole, as the type is not necessarily preserved by the sum of spinors \cite{4}. Regarding representations in the sector of $\Sigma_i$, for which the type is not preserved by the sum, the situation is quite unclear so far. 

\section{Dynamics avatar}

The group of transformations regarding type-$i$ spinors may be faced as an additional step towards the continuity of such a sector of $\Sigma$. However, any spinor in this space may be endowed of a dynamics inherited from the dynamics in spacetime. In this section we will investigate the behavior of spinors as elements in $\Sigma$. Let $\bar{\varphi}$ be a one-to-one, linear, and invertible mapping from $\Sigma$ to sections of ${\bf P}_{Spin_{1,3}^e}\times_\tau \mathbb{C}^4$, i. e. 
\begin{eqnarray}
\bar{\varphi} &: \Sigma \rightarrow {\bf P}_{Spin_{1,3}^e}\times_\tau \mathbb{C}^4 \nonumber\\ & \psi \mapsto \bar{\varphi}[\psi]=\Psi(\vec{x},t). \label{var}  
\end{eqnarray} 
We shall restrict ourselves to the subset $U\subset {\bf P}_{Spin_{1,3}^e}\times_\tau \mathbb{C}^4$ such that the spinors $\Psi(\vec{x},t) \in U$ are subjected to the usual dynamics dictated by the Dirac operator $\mathcal{D}$, i. e. $\mathcal{D}\Psi(\vec{x},t)=0$. In addition, we are going to restrict $\bar{\varphi}$ to ${\varphi}=\bar{\varphi}\mid_{\bar{\varphi}^{-1}(U)}$, namely, the domain of ${\varphi}$ shall be the preimage of $U$, denoted by $\bar{\varphi}^{-1}(U)$. In analogy to the Dirac operator $\mathcal{D}$, let $\nabla$ be a ``dynamical'' operator (an automorphism) in $\Sigma$ such that 
\begin{eqnarray}
\nabla: \Sigma &\rightarrow& \Sigma \nonumber\\  \psi&\mapsto& \nabla\psi, \label{nab} 
\end{eqnarray} whose relation with the dynamical operators be simply given by $\mathcal{D}=\varphi\circ\nabla\circ\varphi^{-1}$. As $\varphi^{-1}\circ \varphi=Id_{\Sigma}$ one has $\nabla=\varphi^{-1}\circ \mathcal{D}\circ \varphi$. Notice, in particular, that the algebraic zero resulting from the action of the Dirac operator is mapped into the null spinor in $\Sigma$. In fact, 
\begin{eqnarray}  
\nabla\psi=\varphi^{-1}\circ \mathcal{D}\circ \varphi[\psi]=\varphi^{-1}\circ \mathcal{D}\Psi(\vec{x},t),\label{pre}
\end{eqnarray} and $\mathcal{D}\Psi(\vec{x},t)=0$ yields $\nabla\psi=0_{\Sigma}$. That is the alluded dynamical avatar which, despite have been straightforwardly obtained, leads to interesting consequences.  

For free fermionic particles in the spacetime, the Dirac operator is usually expressed as $\mathcal{D}=i\gamma^{\mu}\partial_{\mu}-m\mathbb{I}$, where $\gamma^\mu$ are the Dirac matrices and $m$ the mass parameter. Therefore
\begin{eqnarray}
\nabla\psi=\varphi^{-1}\circ (i\gamma^{\mu}\partial_{\mu}-m\mathbb{I})\Psi(\vec{x},t), \label{indo}
\end{eqnarray} which, by means of the map linearity, leads to 
\begin{eqnarray}
i\varphi^{-1}\circ (\gamma^\mu\partial_{\mu}\Psi(\vec{x},t))-m\psi=0_\Sigma. \label{nov}
\end{eqnarray} At first sight, one might speculate that the matrix representation of $\varphi$ commutes with gamma matrices. However, it would imply that $\varphi$ is proportional to the identity \cite{3}. This scenario is too restrictive. Here we will require something less limiting, by demanding the commutation of $\varphi$ only with $\gamma_0$. This requirement will be useful in what follows. Let us denote, then, the pullback of the spinor $\partial_t\Psi(\vec{x},t)$ by $\delta_t\psi:=\varphi^{-1}\circ (\partial_t\Psi(\vec{x},t))=\varphi^{-1}\circ \partial_t \circ \varphi[\psi]$, and then write 
\begin{eqnarray}
i\gamma^0\delta_t\psi+i\varphi^{-1}\circ (\vec{\gamma}\cdot \partial_{\vec{x}}\Psi(\vec{x},t))-m\psi=0_{\Sigma}.\label{primeiro}
\end{eqnarray}

Eq. (\ref{primeiro}) performs a shadow, so to speak, of the spacetime dynamics respected by the physical spinor. It may be applied to every sector of $\varphi^{-1}(U)\subset \Sigma$ and in this space as a whole. However, it is not completely clear so far which connections may be reached inside the spinor space, see for instance \cite{jmp}. Therefore we will assume, in a first moment, a conservative approach adopting the physically sound particularization that the spinor type is not changed by the dynamics and study its consequences for each $\Sigma_i$ separately.    

The spinors belonging to $\Sigma_i$ are called physical, in the sense that they satisfy the FPK identities. The attribute ``physical" in dealing with spinors, however, must be used with a great care. In fact, a spinor alone describing a fermion cannot be detected. Its dual -- and the correspondent theory -- must be taken into account. Despite of these important matters, if the elements of $\Sigma_i$ are representatives of physical states, then they have to be conserved. Consider a macroscopically dense set of spinors in $\mathcal{F}\subset\Sigma_i$ and suppose that the surface $\partial\mathcal{F}$ is orientable. A conservation law will encounter an analogue within $\mathcal{F}$. Hence, being the density $\rho$ of spinor states in this region characterized by $\rho(\psi,t)$ one may be able to write 
\begin{equation}     
\frac{\partial \rho}{\partial t}=-\frac{\delta (\rho \delta_t \psi)}{\delta \psi}. \label{cons}
\end{equation} Some considerations concerning Eq. (\ref{cons}) are in order. As to represent a conservation law, $\delta_t \psi$ denotes a generalized velocity in $\mathcal{F}$ leading, then, the term $\rho\delta_t \psi$ to express a current of states. Therefore, a decreasing [increasing] in the density of states is taken due to an output [input] current. Besides, the functional derivative present in Eq. (\ref{cons}) may be taken in exact same footing as its counterpart in classical and quantum field theory. We are now in position to assert the following result, in close analogy to the Statistical Mechanics Liouville theorem for physical states in the phase space \cite{pat}.

\vspace{.2cm}
{\bf Proposition:} {\it Massless spinors representing conserved states in $\mathcal{F}\subset \Sigma_i$, with orientable $\partial\mathcal{F}$, behave as an incompressible fluid.} 
\vspace{.2cm} 

{\bf Proof:} The time variation of the representative density reads
\begin{equation}
\frac{d\rho(\psi,t)}{dt}=\frac{\delta \rho(\psi,t)}{\delta \psi}\delta_t\psi+\frac{\partial \rho(\psi,t)}{\partial t}
\end{equation} and taking (\ref{cons}) into account yields 
\begin{equation}
\frac{d\rho(\psi,t)}{dt}=-\rho \frac{\delta (\delta_t \psi)}{\delta \psi}.\label{dois}
\end{equation} The equation governing the behavior in the spinor space, (\ref{primeiro}), for massless spinors may be recast into the form 
\begin{equation}      
\mathbb{I}\delta_t \psi=-\gamma^0\varphi^{-1}\circ (\vec{\gamma}\cdot\partial_{\vec{x}}\Psi(\vec{x},t)).\label{sei} 
\end{equation}  From Eqs. (\ref{dois}) and (\ref{sei}) it is fairly simple to see that 
\begin{equation}
\frac{d(\mathbb{I} \rho(\psi,t))}{dt}=0,\label{ies}
\end{equation} culminating in four identical equations satisfied by a constant density. \hspace{.2cm}  $\Box$\medbreak

The massive case is just inconsistent. We are currently investigating this case, for which, we speculate, none conservative equation analogue can be stated, but have not a satisfactory interpretation for that so far. Before concluding this section, we would like to contrast our results with the case concerning exotic spinors.     

It is well known that when the base manifold, $M$, is not simply connected there is not only one spinorial structure \cite{exo1}. This fact is traduced by the non triviality of the (first) cohomology group $H^1(M, \mathbb{Z}_2)$. The non trivial topology is then reflected in the dynamics \cite{exo2,exo21}, by means of an additional term in the Dirac operator now reading $\tilde{\mathcal{D}}=i\gamma^{\mu}\partial_{\mu}+i\gamma^\mu\partial_\mu\theta(\vec{x},t)-m\mathbb{I}=\mathcal{D}+i\gamma^\mu\partial_\mu\theta(\vec{x},t)$, where $\theta(\vec{x},t)$ is a real scalar function (for mathematical details see, for instance, \cite{exo3}). Concerning the topics approached in Sect. \ref{II}, there are very few substantial differences in dealing with exotic spinors instead of usual spinors. As a matter of fact, it is still possible to define a spinor space for exotic spinors and, as before, we also will have a categorization of the spinorial space into types. The unique novelty is that there are three more possible types of spinors, but the results of the previous section remain valid. A noteworthy difference occurs in the appreciation of the proposition above to exotic spinors. Denoting exotic spinors in the exotic spinorial space ($\tilde{\Sigma}$) by $\tilde{\psi}$ (and its spacetime counterpart by $\tilde{\Psi}(\vec{x},t)$), it is fairly direct to see that the analogue of Eq. (\ref{primeiro}) for the case at hands reads    
\begin{eqnarray}
i\gamma^0\delta_t\tilde{\psi}-m\tilde{\psi}+i\gamma^0\dot{\theta}(\vec{x},t)\tilde{\psi}+i\varphi^{-1}\circ \Big(\vec{\gamma}\cdot\{\partial_{\vec{x}}\tilde{\Psi}(\vec{x},t)+\partial_{\vec{x}}\theta(\vec{x},t)\tilde{\Psi}(\vec{x},t)\}\Big)=0_{\tilde{\Sigma}},\label{penult}
\end{eqnarray} where $\dot{\theta}=\partial_t\theta$. In this vein, massless exotic spinors shall obey 
\begin{eqnarray}
\mathbb{I}\delta_t\tilde{\psi}=-\mathbb{I}\dot{\theta}(\vec{x},t)\tilde{\psi}-\gamma^0\varphi^{-1}\circ \Big(\vec{\gamma}\cdot\{\partial_{\vec{x}}\tilde{\Psi}(\vec{x},t)+\partial_{\vec{x}}\theta(\vec{x},t)\tilde{\Psi}(\vec{x},t)\}\Big).\label{ult}
\end{eqnarray} Though the form of Eq. (\ref{ult}) is not particularly clear from the physical point of view, the investigation of the exotic spinors behavior in $\tilde{\Sigma}$ leads to the fact that the spinorial density, provided conservation, is given in terms of the exotic additional term $\rho(\psi,t)=\rho_0\exp(\theta(\vec{x},t))$. Of course, $\rho_0$ is constant in such a way that if $\theta=0$ (the usual case of trivial topology) the proposition result is recovered, as expected.

\section{Concluding remarks}

In Sec. III we show the possibility of symmetries transformations in the spinor space as elements of a subgroup of $GL(4,\mathbb{C})$. These symmetries respect the Lounesto classification and so do not accross the spinor type. While relevant results on their own, we would like here to give a comprehensive account on results. In Ref. \cite{nove} it was proposed an interpolation between sectors of a given representation, encompassing spinors satisfying the Heisenberg equation of motion, which could lead to the neutrino oscillation even in the massless case. All these spinors was shown to belong to Lounesto type-1 case \cite{ult}. The results here explored may serve as a first step towards the mathematical investigation of such an interpolation, in the sense that it was conjectured to be performed by an unitary operator \cite{nove} whose action preserves the spinor type \cite{ult}.

In Sec. IV, we explore the interelationship between the dynamics occurring in the spacetime and its reflex in the spinor space. The interplay between spinors, bilinear covariants and hydrodynamics was implemented in Refs. \cite{daRocha:2016bil,Bonora:2015ppa}, in the context of the Lounesto spinor classification. In Ref. \cite{Bonora:2015ppa}
suitable black hole backgrounds were considered, having  a current density that interpolates
between a timelike Killing vector field at the spatial infinity and the null Killing vector
field on the black hole event horizon. This current density was identified to a spinor
fluid flow. In Ref. \cite{daRocha:2016bil}, flag-dipole spinors, satisfying the Dirac equation in another black hole background was shown to induce an underlying fluid flow structure in the background  spacetime. These two results are quite particular, relating fluid mechanics to the 
Lounesto classification. On the other hand, the results in the Proposition here presented are universal, relating the dynamics of certain spinors with the equations of motion of incompressible fluids. The investigation of this result in the context of exotic spinors was presented. It was shown that unusual topology in the spacetime leads to a modification in the spinor space dynamics. While some modification is generically expected, since the connection is changed, we emphasize that the dynamical interplay was strong enough to reveal that unusual topology forbids the perfect fluid behavior. We are currently investigating additional developments of this interplay, as well its limitations.


\section*{Acknowledgments}
JMHS thanks to CNPq (Grant No. 303561/2018-1) for partial financial support. RdR is grateful to FAPESP (Grant No. 2017/18897-8), to CNPq (Grants No. 406134/2018-9 and No. 303293/2015-2) and to HECAP - ICTP, Trieste, for partial financial support, and this last one also for the hospitality. RTC is grateful to CAPES and UNESP $|$ Guaratingueta post-graduation program for financial support.

\end{document}